\documentclass[journal]{IEEEtran}

\usepackage{amssymb}
\usepackage{amsfonts}
\usepackage{amsmath}
\usepackage{graphicx}
\usepackage{cite}
\usepackage{graphics} 

\newtheorem{thm}{Theorem}
\newtheorem{lem}{Lemma}

\newtheorem{cor}{Corollary}

\newtheorem{exmp}{Example}
\newtheorem{rem}{Remark}

\def\startprf{{\emph{Proof}. }}
\def\endprf{{$\hfill\square$\\}}

\begin{document}

\title{Stabilization of Linear Systems Across a Time-Varying AWGN Fading Channel}

\author{Lanlan Su, Vijay Gupta and Graziano Chesi\thanks{L. Su and G. Chesi are with the Department of Electrical and Electronic
Engineering, The University of Hong Kong.
V. Gupta is with Department of Electrical Engineering, University
of Notre Dame. (Email: llsu@eee.hku.hk, vgupta2@nd.edu, chesi@eee.hku.hk).}}



\maketitle

\begin{abstract}
This technical note investigates the minimum average transmit power required for mean-square
stabilization of a discrete-time linear process across a time-varying
additive white Gaussian noise (AWGN) fading channel that is presented between the  sensor  and the controller. We assume channel state information at both the transmitter and the receiver, and allow the transmit power to vary with the channel state to obtain
 the minimum required average transmit power via optimal power
adaptation.  We consider both the case of independent and identically distributed fading 
and fading subject to a  Markov chain. Based on the proposed necessary and sufficient conditions for mean-square stabilization, we show that the minimum average
transmit power to ensure stabilizability can be obtained by solving  a
geometric program. \end{abstract}


\section{Introduction}
The interaction between control and communication plays a
crucial role in networked control systems, which have received
considerable attention across the past decade. In particular,
the problem of feedback stabilization in the presence of
communication channel models has been widely studied in
the literature (see, e.g., \cite{cetinkaya2016networked} which deals with the packet loss
channel, \cite{kumar2014stabilizability, freudenberg2010stabilization} which are concerned with an additive
white Gaussian noise (AWGN) channel, 
\cite{xu2017mean, xu2017mean2} which consider a fading channel and \cite{minero2009data,minero2013stabilization} wherein a digital
communication link with time-variant rate is considered).

In this work, we are interested in control across a time-varying block-fading channel subject to AWGN with channel state information at both the transmitter and the receiver. A fading channel is defined by an input-output relation $y=gx+w$ where the channel gain $g$ is time-varying used to characterize the various processes encountered by transmitted waves on their way from the transmitter and receiver antennas, and the  noise $w$ is often assumed to be AWGN used to describe the additive noise generated within the receiver. The term block fading refers to the case when the gain $g$ remains constant during blocks of time and varies from one block to the next. This is a common model used for the wireless channel \cite{quevedo2013state,minero2009data}. If the fading channel varies sufficiently slowly, the value of the gain during a particular block can be assumed to be known causally to both the transmitter and the receiver (see, e.g., \cite{jayaweera2003capacity,el2011network} and the references therein). We make the same assumption in this work. There are relatively few works that consider  control across a fading channel. \cite{minero2009data, minero2013stabilization}    use time-varying  channels; however, the time-variation considered is in the number of bits that can be transmitted in a noiseless fashion (the so-called digital noiseless channel model) rather than in the gain (and hence the signal to noise ratio). \cite{qi2017control,su2017control} consider control over a fading channel where the channel gain varies in an independent and identically distributed (i.i.d.) manner. However, they assume that no noise is present and the value of the gain $g$ is unknown to either the receiver or the transmitter. \cite{xu2017mean,xu2017mean2} consider the presence of noise; however, they assume a fast-fading model in which the gain changes at every time step and is  fed back to the transmitter through a noiseless feedback channel with one-step delay.

Motivated by
the fact that typically the channel gain  varies at a much longer time scale
than symbol transmission time \cite{el2011network}, we consider a block-fading channel model in this work. Specifically, we investigate the minimum average transmit power required for mean-square stabilization of a discrete-time linear process 
across a time-varying AWGN block-fading channel that is presented from the  sensor  to  the controller. We assume channel state information at both the transmitter and the receiver, and allow the transmit power to vary with the channel state.   Both the case when the channel gain varies in an i.i.d. fashion among blocks and when it varies according to a Markov chain are considered. In either case, we provide a tight characterization of the minimum average transmit power required for stabilization. It is worth mentioning that generalization
of results from fast-fading channel to block-fading channel is not trivial from
the perspective of theoretical proof.  The main contribution of this note is showing that by
allowing power adaptation, the minimum transmit
power to ensure stabilizability can be obtained by solving a geometric program.
This reveals an interesting difference from the water-filling interpretation of distributing  transmit power in time for achieving channel capacity \cite{goldsmith1997capacity}, which arises from the fact that the objective here is stabilization rather than achieving capacity.  To the best of our knowledge, this has
never been proposed in the literature. 

\textbf{Notation:}
For a continuous random variable $X$, the differential entropy of $X$ is denoted by $h(X)$. The mutual information between two
continuous random variables $X,Y$ is denoted as $I(X;Y)$.
The  expectation operator is denoted by $\mathbb{E}[\cdot]$, and the random variable over which the expectation is taken is usually clear from the context. The notation $\rho (A)$
denotes the spectral radius of $A$. The notation $\text{log}$ denotes the natural logarithm.

\section{Problem Formulation}
\begin{figure}
\center \includegraphics[scale=0.22]{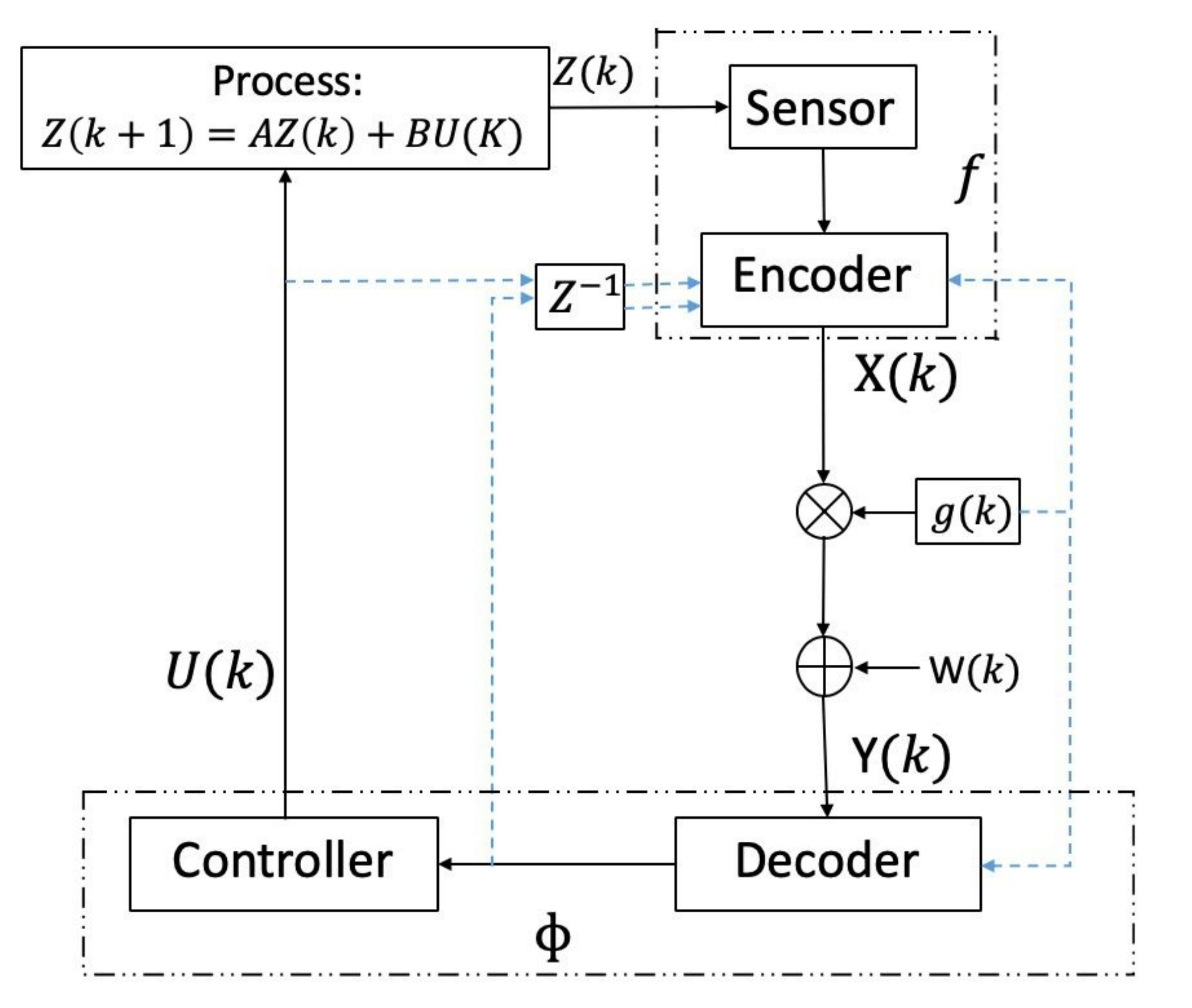}
\center \caption{Feedback control across a time-varying AWGN fading channel }\label{fig:0}
\end{figure}
Consider the closed-loop system depicted in Figure \ref{fig:0}.
The linear time-invariant process is represented as
\begin{equation}
Z(k+1)=AZ(k)+BU(k),\label{eq:vector_system}
\end{equation}
where $Z(k)\in\mathbb{R}^{l}$ is the state, and $U(k)\in\mathbb{R}$
is the input. For simplicity, we assume that the  matrix
$A\in\mathbb{R}^{l\times l}$ is in the Jordan form with only
unstable modes, and the pair $(A,B)$ is controllable. Each component of the
initial condition, $Z^{i}(0)$, is randomly distributed with mean $\mu_{Z^{i}(0)}$  and variance
$\sigma_{Z^{i}(0)}^{2}$.

 As shown in Figure \ref{fig:0}, the input and the output of the channel from encoder to decoder at time $k$ are denoted as $X(k)$ and $Y(k)$ with
$Y(k)=g(k)X(k)+W(k)$ where $g(k)$ represents the attenuation
gain due to the fading and $W(k)$ is a zero-mean Gaussian white
noise process with variance $N$. The gain $g(k)$ is allowed to
be time-varying across time blocks of length $n$ $(n>1)$. In other words, $g(k)$
remains constant in blocks of time $\{0,\ldots,n-1\}$,$\{n,\ldots,2n-1\}$, ..., and
 varies among these blocks. Assume that $g(k)$ takes value in a finite
set $\mathcal{S}=\left\{ g_{1},g_{2},\ldots,g_{m}\right\}$ with $\mathcal{S}\neq \left\{ 0\right\}$. During
the $j$-th block, $k\in \{ jn,\ldots,(j+1)n-1 \}$, we denote  $g(k)$ as $g_{\sigma_{(j)}}$where
$\sigma(j)$ is a switching signal taking values in the set $\left\{ 1,\ldots,m\right\} $,
and we say that the channel state is $s$ if $\sigma(j)=s$. It is assumed that $\sigma(j)$ is switching either according to an i.i.d.
process or as governed by a Markov chain across blocks. We make the additional simplifying assumption
that $\frac{n}{l}\in\mathbb{N}$.

At  every time step $k$, an encoder at the sensor calculates the channel input  $X(k)$ by a function $f$. We assume that the encoder has access to one step delayed control input $U(k-1)$. This does not necessarily require the controller to send directly its output to the encoder since $U(k-1)$ can be calculated via equation \eqref{eq:vector_system} from the  observation of $Z(k-1)$ and $Z(k)$, i.e., $U(k-1)=|Z(k)-AZ(k-1)|/|B|$. It is also assumed that the encoder has access to one-step delayed decoder output. This can be realized by a perfect feedback channel from decoder to encoder, or by exploiting a smart controller that can send additional extractable information to the actuator. The constraint imposed on the encoder is through the average transmit power $\text{lim}_{t\rightarrow\infty}\frac{1}{t}\sum_{k=0}^{t-1}\mathbb{E}\left[X^{2}(k)\right]\leq P$. Since the encoder has channel state information, we allow its transmit power to be adapted according to the channel state. Denote $P_{\sigma(j)}=\frac{1}{n}\sum_{k=jn}^{(j+1)n-1}\mathbb{E}\left[X^{2}(k)\right]$, and then the constraint becomes $\text{lim}_{T\rightarrow\infty}\frac{1}{T}\sum_{j=0}^{T-1}P_{\sigma(j)}\leq P$. The decoder is collocated with the controller and calculates the control input by a function $\phi$. Note that  $f$ and $\phi$ are allowed to be of any causal functions of  all the available information till the time step $k$.

Recall that the process \eqref{eq:vector_system} is said to be mean-square stabilizable if there exist an encoder-decoder pair and a controller such that $\text{lim}_{k\rightarrow\infty}\mathbb{E}\left[Z(k)Z(k)^{T}\right]=0$
for any initial condition $Z(0)$.
The question we raise is what is the minimum average transmit power $P^{*}$ that is required to
stabilize the process \eqref{eq:vector_system} in the mean square
sense with the problem formulation stated above and any design of the functions $f$ and $\phi$?

We will use the following result that converts the control problem to an estimation problem.

\begin{lem}
\label{lem:1}\cite{kumar2014stabilizability} Let $\hat{Z}_{0}(k)$
be the estimate of the initial state $Z(0)$ at the $k$-th time step calculated by some decoder $\mathcal{D}$ using the information it has access to. Denote the estimate error as $\epsilon(k)\triangleq \hat{Z}_{0}(k)-Z(0)$.
If 
\begin{equation}
\mathbb{E}\left[\epsilon(k)\right]=0,\label{eq:Lemma1 con1}
\end{equation}\begin{equation}
\lim_{k\rightarrow\infty}A^{k}\mathbb{E}\left[\epsilon(k)\epsilon^{T}(k)\right]\left(A^{T}\right)^{k}=0,\label{eq:Lemma 1 con2}
\end{equation}
then the process \eqref{eq:vector_system} can be stabilized in the
mean square sense by the controller $U(k)=K\bar{Z}(k)$ with $K$ chosen such that $A+BK$
is Schur stable and $\bar{Z}(k)=A^{k}\hat{Z}_{0}(k)+\sum_{j=1}^{k}A^{k-j}BU(j-1)$. The function $\phi$ is hence given as $\phi\triangleq A^k\mathcal{D}(k,g(0),\ldots,g(k),Y(0),\ldots,Y(k))+\sum_{j=1}^{k}A^{k-j}BU(j-1)$.
\end{lem}

Note that the condition \eqref{eq:Lemma 1 con2} is merely an asymptotic condition. Thus, considering the subsequence $k=jn-1,j=1,2,\ldots$, the condition \eqref{eq:Lemma 1 con2}
can be equivalently written as 
\begin{equation}
\textstyle
\lim_{j\rightarrow\infty}A^{jn-1}\mathbb{E}\left[\epsilon(jn-1)\epsilon^{T}(jn-1)\right]\left(A^{T}\right)^{jn-1}=0.\label{eq:Lemma_cond2_2}
\end{equation}

For brevity, we  use the notation $X_{k}$ and
$Y_{k}$ to denote  $X(k)$ and $Y(k)$.
Let $\sigma_0^j$ and $Y_{i}^{k}$ represent the sequence of channel state from initial block to the $j$-th block, i.e., $\sigma_0^k\triangleq\{\sigma(0),\ldots,\sigma(j)\}$ and  the collection of the observations
of the decoder from time step $i$ to $k$, i.e.,  $Y_{i}^{k}\triangleq \left\{ Y(i),\ldots,Y(k)\right\}$, and we use $\bar{\sigma}_0^j$ and $y_i^k$ to denote their realizations.

\section{ The i.i.d case\label{sec:3}}
In this section, we consider the case where the channel state $\sigma(j)$
is switching according to an i.i.d. process across the blocks.  We start with the case when $Z(k)\in \mathbb{R}$  and then consider the more general vector case. 
\subsection{Scalar systems}
Consider the case when $Z(k)\in \mathbb{R}$. The process with $|\lambda|>1$ is described as
\begin{equation}
Z(k+1)=\lambda Z(k)+U(k).\label{eq:scalar system}
\end{equation}
Let $\mathbb{E}_{\sigma(j)}\left[F(\sigma(j))\right]=\sum_{s=1}^{m}\text{Pr}(\sigma(j)=s)F(s)$.
\begin{thm}
\label{thm:1}Given $P_1,\ldots,P_m$, the process \eqref{eq:scalar system}
is mean-square stabilizable across the time-varying AWGN fading channel
if and only if
\begin{equation}
\text{log}|\lambda|<-\frac{1}{2n}\text{log}\;\mathbb{E}_{\sigma(j)}\left[\left(\frac{N}{g_{\sigma(j)}^{2}P_{\sigma(j)}+N}\right)^{n}\right].\label{eq:nec_suf_iid_scalar}
\end{equation}
\end{thm}
\startprf
''$\Leftarrow$'' We prove the sufficiency by applying the Elias scheme  in the context of feedback control that will generate
the initial state estimate $\hat{Z}_{0}(k)$ with error $\epsilon(k)$
satisfying \eqref{eq:Lemma1 con1} and \eqref{eq:Lemma_cond2_2}.
Let us denote the variance of $\epsilon(k)$ at the end of the $j$-th block  as $\alpha(j)=\mathbb{E}\left[\epsilon^{2}((j+1)n-1)\right]$. 

During the initial block, $k\in\left\{ 0,1,\ldots,n-1\right\} $,
the channel state is $\sigma(0)$. At time $k=0$, the encoder transmits
$X(0)=\sqrt{\frac{P_{\sigma(0)}}{\sigma_{Z(0)}^{2}}}\left(Z(0)-\mu_{Z(0)}\right)$, the decoder
computes the unbiased estimate  $\hat{Z}_{0}(0)=\frac{1}{g_{\sigma(0)}}\sqrt{\frac{\sigma_{Z(0)}^{2}}{P_{\sigma(0)}}}Y(0)+\mu_{Z(0)}$,
and therefore the variance of the estimate error is $\mathbb{E}\left[\epsilon^{2}(0)\right]=\frac{\sigma_{Z(0)}^{2}N}{g_{\sigma(0)}^{2}P_{\sigma(0)}}$.
At each time $k\geq1$, the encoder transmits $X(k)=\sqrt{\frac{P_{\sigma(0)}}{\mathbb{E}[\epsilon^{2}(k-1)]}}\epsilon(k-1)$
and the decoder computes the new unbiased estimate $\hat{Z}_{0}(k)$
based on $Y(k)$ and $\hat{Z}_{0}(k-1)$ by linear minimum mean square
error (MMSE) estimation, i.e., 
\begin{equation}
\hat{Z}_{0}(k)=\hat{Z}_{0}(k-1)-\frac{\mathbb{E}\left[Y(k)\epsilon(k-1)\right]}{\mathbb{E}\left[Y^{2}(k)\right]}Y(k).\label{eq:MMSE}
\end{equation}
Accordingly, the variance of the estimate error at the end of the
initial block is given by 
\[
\alpha(0)=\mathbb{E}_{\sigma(0)}\left[\frac{\sigma_{Z(0)}^{2}N}{g_{\sigma(0)}^{2}P_{\sigma(0)}}\left(\frac{N}{g_{\sigma(0)}^{2}P_{\sigma(0)}+N}\right)^{n-1}\right].
\]

During the $j$-th block, $j\geq1$, the encoder transmits $X(k)=\sqrt{\frac{P_{\sigma(j)}}{\mathbb{E}[\epsilon^{2}(k-1)]}}\epsilon(k-1)$
at each time $k\in\left\{ jn,jn+1,\ldots,(j+1)n-1\right\} $, and
the decoder computes the new estimate $\hat{Z}_{0}(k)$ based on $Y(k)$
and $\hat{Z}_{0}(k-1)$ by the linear MMSE estimation \eqref{eq:MMSE}.
In this way, the recursion of $\alpha(j)$ is given by 
$
\alpha(j)=\alpha(j-1)\mathbb{E}_{\sigma(j)}\left[\left(\frac{N}{g_{\sigma(j)}^{2}P_{\sigma(j)}+N}\right)^{n}\right].
$

It follows from \eqref{eq:nec_suf_iid_scalar} that the condition
\eqref{eq:Lemma_cond2_2} is satisfied. Besides, since $\mathbb{E}\left[\epsilon(0)\right]=0$
and the linear MMSE estimator is unbiased, the condition \eqref{eq:Lemma1 con1}
is satisfied. Therefore, based on Lemma \ref{lem:1}, the process \eqref{eq:scalar system}
is mean-square stabilizable.

''$\Rightarrow$'' The necessity is obtained via similar information-theoretic arguments in \cite{minero2009data,xu2017mean,freudenberg2010stabilization} with the differences caused by the block analog channel. First, let us define the conditional entropy power
of $Z(jn)$ conditional on the event $Y_{0}^{jn-1}=y_{0}^{jn-1}$, $\sigma_0^{j-1}=\bar{\sigma}_0^{j-1}$, and averaged over $Y_{0}^{jn-1},\sigma_0^{j-1}$ as $N(j)\triangleq\frac{1}{2\pi e}\mathbb{E}_{Y_{0}^{jn-1},\sigma_0^{j-1}}\left[e^{2h\left(Z(jn)|y_{0}^{jn-1},\bar{\sigma}_0^{j-1}\right)}\right]$.
It follows from the maximum entropy theorem that $N(j)\leq\mathbb{E}[Z^{2}(jn)].$
Thus, a necessary condition for the mean-square stability of the system is $\lim_{j\rightarrow\infty}N(j)=0$. Next, we have
\[
\begin{array}{l}
\ \ N(j+1)\\
=\frac{1}{2\pi e}\mathbb{E}_{Y_{0}^{\left(j+1\right)n-1},\sigma_0^j}\left[e^{2h\left(Z\left(\left(j+1\right)n\right)|y_{0}^{\left(j+1\right)n-1},\bar{\sigma}_0^j\right)}\right]
\\
\stackrel{(a)}{=}\frac{\lambda^{2n}}{2\pi e}\mathbb{E}_{Y_{0}^{\left(j+1\right)n-1},\sigma_0^j}\left[e^{2h\left(Z\left(jn\right)|y_{0}^{\left(j+1\right)n-1},\bar{\sigma}_0^j\right)}\right]
\\
\stackrel{(b)}{\geq}\frac{\lambda^{2n}}{2\pi e}\mathbb{E}_{Y_{0}^{jn-1},\sigma_0^j}\\
\ \ \left[e^{2\mathbb{E}_{Y_{0}^{\left(j+1\right)n-1},\sigma_0^j|Y_{0}^{jn-1},\sigma_0^j}\left[h\left(Z\left(jn\right)|y_{0}^{\left(j+1\right)n-1},\bar{\sigma}_0^j\right)\right]}\right]
\end{array}
\]
where $(a)$ follows from the facts that $Z((j+1)n)=\lambda^{n}Z(jn)+\sum_{k=jn}^{\left(j+1\right)n-1}\lambda^{\left(j+1\right)n-k-1}U(k)$
and $U(k),k\in \{jn,\ldots,(j+1)n-1\}$ is a function of $Y_{0}^{k}$ and $\sigma_0^j$, $(b)$ is based on the
law of total expectation and Jensen's inequality. 

It can be observed that
\[
\begin{array}{l}
\ \ \mathbb{E}_{Y_{0}^{\left(j+1\right)n-1},\sigma_0^j|Y_{0}^{jn-1},\sigma_0^j}\left[h\left(Z\left(jn\right)|y_{0}^{\left(j+1\right)n-1},\bar{\sigma}_0^j\right)\right]\\
=h\left(Z\left(jn\right)|Y_{jn}^{\left(j+1\right)n-1},y_{0}^{jn-1},\bar{\sigma}_0^j\right)\\
=h\left(Z\left(jn\right)|y_{0}^{jn-1},\bar{\sigma}_0^j\right)-I\left(Z\left(jn\right);Y_{jn}^{\left(j+1\right)n-1}|y_{0}^{jn-1},\bar{\sigma}_0^j\right)
\end{array}
\]
where the first equality follows from the definition of conditional entropy and the second equality follows from the definition of conditional mutual information.
Since
$Y_{jn}^{\left(j+1\right)n-1}=\left\{ Y_{jn},Y_{jn+1},\ldots,Y_{(j+1)n-1}\right\}$, by the chain rule for mutual information, we have 
\[
\begin{array}{l}
\ \ I\left(Z\left(jn\right);Y_{jn}^{\left(j+1\right)n-1}|y_{0}^{jn-1},\bar{\sigma}_0^j\right)
\\
= \sum_{k=jn}^{(j+1)n-1}I\left(Z\left(jn\right);Y_{k}|Y_{jn}^{k-1},y_{0}^{jn-1},\bar{\sigma}_0^j\right).
\end{array}
\]
 Moreover,
for each time step in the $j$-th block, $k\in\left\{ jn,jn+1,\ldots,(j+1)n-1\right\} ,$
the random variable $Z\left(k\right)$ is a function of $Z(jn)$ given $Y_{0}^{k-1}$ and $\sigma_0^j$, and vice versa. This leads to 
\[
\begin{array}{rl}
 & \sum_{k=jn}^{(j+1)n-1}I\left(Z\left(jn\right);Y_{k}|Y_{jn}^{k-1},y_{0}^{jn-1},\bar{\sigma}_0^j\right)\\
=&\sum_{k=jn}^{(j+1)n-1}I\left(Z\left(k\right);Y_{k}|Y_{jn}^{k-1},y_{0}^{jn-1},\bar{\sigma}_0^j\right)\\
\stackrel{(c)}{\leq}&\sum_{k=jn}^{(j+1)n-1}I\left(X_{k};Y_{k}|Y_{jn}^{k-1},y_{0}^{jn-1},\bar{\sigma}_0^j\right)\\
\stackrel{(d)}{\leq}&nC_{\bar{\sigma}(j)},
\end{array}
\]
where $C_{\bar{\sigma}(j)}=\frac{1}{2}\text{ln}\left(1+\frac{g_{\bar{\sigma}(j)}^{2}P_{\bar{\sigma}(j)}}{N}\right)$ is the channel capacity of the AWGN fading channel with gain $g_{\bar{\sigma}(j)}$,  $(c)$ is obtained since $Z(k)\rightarrow X_{k} \rightarrow Y_k$ form a Markov chain, and $(d)$ follows from the definition of  Gaussian channel capacity. In view of independence between $Z\left(jn\right)$ and $\sigma(j)$,
we have $h\left(Z\left(jn\right)|y_{0}^{jn-1},\bar{\sigma}_0^j\right)=h\left(Z\left(jn\right)|y_{0}^{jn-1},\bar{\sigma}_0^{j-1}\right).$
Thus, it can be obtained that
\[
\begin{array}{l}
\ \ \mathbb{E}_{Y_{0}^{\left(j+1\right)n-1},\sigma_0^j|Y_{0}^{jn-1},\sigma_0^j}\left[h\left(Z\left(jn\right)|y_{0}^{\left(j+1\right)n-1},\bar{\sigma}_0^j\right)\right]\\
\geq h\left(Z\left(jn\right)|y_{0}^{jn-1},\bar{\sigma}_0^{j-1}\right)-nC_{\bar{\sigma}(j)}.
\end{array}
\]
Consequently, it follows that 
\[
\begin{array}{l}
 \ \ N(j+1)\\
\geq  \frac{\lambda^{2n}}{2\pi e}\mathbb{E}_{Y_{0}^{jn-1},\sigma_0^j}\left[e^{2h\left(Z\left(jn\right)|y_{0}^{jn-1},\bar{\sigma}_0^{j-1}\right)-2nC_{\bar{\sigma}(j)}}\right]\\
=  \frac{\lambda^{2n}}{2\pi e}\mathbb{E}_{Y_{0}^{jn-1},\sigma_0^{j-1}}\left[e^{2h\left(Z\left(jn\right)|y_{0}^{jn-1},\bar{\sigma}_0^{j-1}\right)}\right]\mathbb{E}_{\sigma(j)}\left[e^{-2nC_{\bar{\sigma}(j)}}\right]\\
=  N(j)\lambda^{2n}\mathbb{E}_{\sigma(j)}\left[\left(\frac{N}{g_{\sigma(j)}^{2}P_{\sigma(j)}+N}\right)^{n}\right].
\end{array}
\]
Hence, we can prove the necessity of condition \eqref{eq:nec_suf_iid_scalar}
via contradiction since if the condition \eqref{eq:nec_suf_iid_scalar}
is violated, then $N(j)$ will not converge to zero.\endprf

\begin{exmp}\label{ex:1}
Consider the scalar process \eqref{eq:scalar system} with $\lambda>1$. The variance of the additive Gaussian noise and the length of the block is given by $N=1$ and $n=20$, and the fading gain subject to an i.i.d. process has two states $g_{1}=1,g_{2}=0.5$. Assume that $\text{Pr}(\sigma(j)=1)=\text{Pr}(\sigma(j)=2)=0.5$. Given $P_1=5,P_2=4.7$, according to Theorem \ref{thm:1}, the maximum $\lambda$ satisfying the condition in \eqref{eq:nec_suf_iid_scalar} is $\lambda_{max}=1.5$, and the average transmit power is $\sum_{s=1}^2 \text{Pr}(\sigma(j)=s)P_s=4.85$.  It will be seen later in Example \ref{ex:2} that the power allocation $P_1=5,P_2=4.7$ is not optimal for minimizing the average  transmit power.
\end{exmp}

\begin{rem}
When $n=1$ and $P_1=\cdots=P_M$, the condition proposed in Theorem \ref{thm:1} coincides with the one provided in  \cite{xu2017mean} which considers a fast-fading channel model. Theorem \ref{thm:1} is an important extension of \cite{xu2017mean} since  under the setting of block-fading model, the transmit power can be adapted dynamically to the channel state. It will be shown in Section \ref{sec:5} that the minimum average transmit power can be significantly reduced when optimal power adaptation is adopted. \end{rem}

\subsection{Vector systems \label{sub:Vector-systems iid}}

In this subsection, we consider the vector plant \eqref{eq:vector_system}
with a specific time-division multiple access (TDMA) scheduling scheme to allocate channel resources
among the subsystems.

Let $\lambda_{1},\lambda_{2}\ldots,\lambda_{l}$ denote the eigenvalues
of $A$ counting algebraic multiplicity. Since $A$ has a Jordan form,
each component of the initial state, denoted as $Z^{i}(0)$, increases
with rate dominated by an eigenvalue $\lambda_{i}$. We now present
a TDMA scheduling strategy for the vector plant \eqref{eq:vector_system}.
We divide every block into $l$ equal-length time slots, and allocate
each time slot of length $\frac{n}{l}$ periodically to each subsystem.
For blocks with channel state $s,s\in\{1,2,\ldots,m\},$ we assign
different  power, $P_{s,1},P_{s,2},\ldots,P_{s,l}$, to the time slots
allocated to subsystems associated with $\lambda_{1},\lambda_{2},\ldots,\lambda_{l}$.
For each $s\in{1,\ldots,m}$, we restrict the set $\{P_{s,1},P_{s,2},\ldots,P_{s,l}\}$ to satisfy
\begin{equation}
|\lambda_{1}|^{2}\left(\textstyle\frac{N}{g_{s}^{2}P_{s,1}+N}\right)^{\frac{1}{l}}=\cdots=|\lambda_{l}|^{2}\left(\textstyle\frac{N}{g_{s}^{2}P_{s,l}+N}\right)^{\frac{1}{l}}.\label{eq:constraint}
\end{equation}

\begin{thm}
\label{thm:2}Given $P_{s,1},\ldots,P_{s,l},s={1,\ldots,m}$ and under the
TDMA scheduling scheme with the constraint \eqref{eq:constraint},
the process \eqref{eq:vector_system} is mean-square stabilizable if and only if 
\begin{equation}
\sum_{i=1}^{l}\text{log}|\lambda_{i}|<-\frac{l}{2n}\text{log}\left[\mathbb{E}_{\sigma(j)}\prod_{i=1}^{l}\left(\frac{N}{g_{\sigma(j)}^{2}P_{\sigma(j),i}+N}\right)^{\frac{n}{l^{2}}}\right].\label{eq:nec_suf_vector}
\end{equation}
\end{thm}
\startprf
''$\Leftarrow$'' We prove the sufficiency by showing that if the condition
\eqref{eq:nec_suf_vector} holds under the described TDMA scheduling
scheme with constraint \eqref{eq:constraint}, the condition \eqref{eq:Lemma1 con1}
and \eqref{eq:Lemma_cond2_2} can be satisfied by adopting a similar
encoder-decoder pair to the one used in the scalar case. Specifically, during
the $i$-th time slot of blocks with channel state $s$, the encoder
is scheduled to transmit with  power $P_{s,i}$ suitable information
of the initial state $Z^{i}(0)$ or the error $\epsilon_{i}(k-1)=\hat{Z}_{0}^{i}(k-1)-Z^{i}(0)$,
and the decoder is designed to update the $i$-th component of $\hat{Z}_{0}(k)$
while keeping the other components unchanged, i.e., $\hat{Z}_{0}(k)=\left[\hat{Z}_{0}^{1}(k-1),\ldots,\hat{Z}_{0}^{i}(k),\ldots,\hat{Z}_{0}^{l}(k-1)\right]$,
with initial value $\hat{Z}_{0}(-1)=0$. The controller is chosen according
to Lemma \ref{lem:1}. First, from \eqref{eq:nec_suf_vector}, we
have $\mathbb{E}_{\sigma(j)}\left[\prod_{i=1}^{l}|\lambda_{i}|^\frac{2n}{l}\left(\frac{N}{g_{\sigma(j)}^{2}P_{\sigma(j),i}+N}\right)^{\frac{n}{l^{2}}}\right]<1$,
and it follows from \eqref{eq:constraint} that $\mathbb{E}_{\sigma(j)}\left[|\lambda_{i}|^{2n}\left(\frac{N}{g_{\sigma(j)}^{2}P_{\sigma(j),i}+N}\right)^{\frac{n}{l}}\right]<1$
for all $i\in\{1,\ldots,l\}$. Since the number of channel use for
each state component is $\frac{n}{l}$ in every block, it can be inferred
based on the proof of Theorem \ref{thm:1} that $\text{lim}_{j\rightarrow\infty}|\lambda_{i}|{}^{2jn-2}\mathbb{E}\left[\epsilon_{i}^{2}(jn-1)\right]=0$
for all $i\in\{1,\ldots,l\}$. Second, let us denote the matrix $M(j)=A^{jn-1}\mathbb{E}\left[\epsilon(jn-1)\epsilon^{T}(jn-1)\right](A^{T})^{jn-1}$.
It follows that the condition \eqref{eq:Lemma_cond2_2} holds if all
the diagonal entries of $M(j)$ converge to $0$ as $j\rightarrow\infty$.
For the case where $A$ has a diagonal structure, we have that $\text{lim}_{j\rightarrow\infty}\left[M(j)\right]_{i,i}=\text{lim}_{j\rightarrow\infty}\lambda_{i}^{2jn-2}\mathbb{E}\left[\epsilon_{i}^{2}(jn-1)\right]=0$
for $i\in\{1,\ldots,l\}$. For the general case where $A$ has a general Jordan
form, all diagonal entries of $M(j)$ can be proved converging to
0 using similar arguments in \cite{kumar2014stabilizability}. Besides,
since $\mathbb{E}\left[\epsilon(0)\right]=0$ and the linear MMSE
estimator is unbiased, the condition \eqref{eq:Lemma1 con1} is satisfied.
Therefore, we can conclude based on Lemma \ref{lem:1} that the process
\eqref{eq:vector_system} can be stabilized in the mean square sense.

''$\Rightarrow$'' The necessity can be validated by similar information-theoretic
reasoning in the proof of Theorem \ref{thm:1} with the following
modifications: 1) the conditional entropy power for $Z(jn)\in\mathbb{R}^{l}$
is modified as $N(j)=\frac{1}{2\pi e}\mathbb{E}_{Y_{0}^{jn-1},\sigma_0^{j-1}}\left[e^{\frac{2}{l}h\left(Z(jn)|y_{0}^{jn-1},\bar{\sigma}_0^{j-1}\right)}\right]$, and $N(j)\leq\frac{1}{l}\mathbb{E}\left[|Z(jn)|^{2}\right]$, see \cite{freudenberg2010stabilization} for details; 2) the term
$\lambda^{2n}$ in  $(a)$ and $(b)$ should be replaced by $|\text{det}(A)|^{\frac{2n}{l}}$;
3) the term $nC_{\bar{\sigma}(j)}$ in $(d)$ is equal to $\sum_{i=1}^{l}\frac{n}{2l}\text{ln}\left(1+\frac{g_{\bar{\sigma}(j)}^{2}P_{\bar{\sigma}(j),i}}{N}\right)$. 
\endprf

\begin{rem}
When $l=1$, the condition \eqref{eq:nec_suf_vector} reduces to \eqref{eq:nec_suf_iid_scalar}. It should be mentioned that such a scheduling scheme is not optimal
in minimizing the average power $P$ required to stabilize the process
\eqref{eq:vector_system}. The reason lies in the fact that the channel
capacity over a block with channel state $s$ using a constant  power
$P_{s}$, $\frac{1}{2}n\text{log}\left(1+\frac{g_{s}^{2}P_{s}}{N}\right)$,
is larger than or equal to that with different  power $P_{s,1},P_{s,2},\ldots,P_{s,l}$
for each time slot, $\sum_{i=1}^{l}\frac{n}{2l}\text{log}\left(1+\frac{g_{s}^{2}P_{s,i}}{N}\right)$,
subject to $P_{s}=\frac{1}{l}\sum_{i=1}^{l}P_{s,i}$, which can be
proved by the AM-GM inequality. Hence, it can be implied from \eqref{eq:constraint} 
that when the distribution of $\{|\lambda_{1}|,\ldots,|\lambda_{l}|\}$
is more centralized, the proposed TDMA scheduling scheme becomes less conservative, and it is nonconservative when $|\lambda_{1}|=\cdots=|\lambda_{l}|$. 
\end{rem}

\section{The Markov chain case\label{sec:4}}
In this section, we generalize the results in Section \ref{sec:3}
to the scenario where the channel state varies according to a Markov
chain. It is assumed that the Markov chain with finite states $\sigma(j)\in\{1,2,\ldots,m\}$
is irreducible with all its states positive recurrent and has a stationary
transition matrix $Q=[q_{ij}]$. We start with the case when $Z(k)\in \mathbb{R}$.

\begin{cor}
\label{cor:1} Given $P_1,\ldots,P_m$, the process \eqref{eq:scalar system}
is mean-square stabilizable across the  AWGN fading channel subject
to the Markov chain $\sigma(j)$ if and only if
\begin{equation}
\lambda^{2n}\rho\left(Q^{T}D\right)<1\label{necsufMarkov}
\end{equation}
where $D=\text{diag}\left(\left(\frac{N}{g_{1}^{2}P_{1}+N}\right)^{n},\ldots,\left(\frac{N}{g_{m}^{2}P_{m}+N}\right)^{n}\right).$\end{cor}
\startprf
''$\Leftarrow$'' We prove the sufficiency by adopting the same coding
scheme described in Theorem \ref{thm:1}.
Since the estimate $\hat{Z}_{0}(k)$ at each time step is unbiased,
the condition in \eqref{eq:Lemma1 con1} is satisfied. Denote the
variance of $\epsilon(k)$ at the end of the $j$-th block given
the channel state as $\alpha(j)=\mathbb{E}\left[\epsilon^{2}((j+1)n-1)|\sigma_0^j\right]$.
It follows from the proof of Theorem \ref{thm:1} that 
$
\alpha(0)=\frac{\sigma_{Z(0)}^{2}N}{g_{\sigma(0)}^{2}P_{\sigma(0)}}\left(\frac{N}{g_{\sigma(0)}^{2}P_{\sigma(0)}+N}\right)^{n-1},
$
and the recursive equation for $\alpha(j)$ is given by 
$
\alpha(j)=\alpha(j-1)\left(\frac{N}{g_{\sigma(j)}^{2}P_{\sigma(j)}+N}\right)^{n}.
$
Next, define a new variable $S(j)=\lambda^{jn-1}\sqrt{\alpha(j-1)}$,
and it follows that $S(j+1)=\lambda^{n}\left(\frac{N}{g_{\sigma(j)}^{2}P_{\sigma(j)}+N}\right)^{\frac{n}{2}}S(j)$
where the sequence $\sigma(j),\ j=1,2,\ldots$ is a Markov chain.
It can be observed by following Lemma \ref{lem:1} that if $S(j)$
is mean-square stable, i.e., $\lim_{j\rightarrow\infty}\lambda^{2(jn-1)}\mathbb{E}\left[\alpha(j-1)\right]=0$
where the expectation is taken over the Markov chain $\sigma(j)$,
then the process \eqref{eq:scalar system} is stabilizable in the mean
square sense. Moreover, based on the results on stability of discrete-time
Markov jump linear system \cite{costa1993stability}, it follows that
$S(j)$ is mean-square stable if and only if the condition \eqref{necsufMarkov}
holds. Consequently, we can conclude that if the condition \eqref{necsufMarkov}
holds, then the process \eqref{eq:scalar system} is stabilizable.

''$\Rightarrow$'' Let us denote the conditional entropy power of $Z(jn)$ conditional on the event $Y_0^{jn-1}=y_0^{jn-1},\sigma_0^{j-1}=\bar{\sigma}_0^{j-1}$, and averaged only over $Y_0^{jn-1}$ as $N(j)\triangleq\frac{1}{2\pi e}\mathbb{E}_{Y_{0}^{jn-1}}\left[e^{2h\left(Z(jn)|y_{0}^{jn-1},\bar{\sigma}_0^{j-1}\right)}\right]$. It can be shown following similar arguments of necessity proof of Theorem \ref{thm:1} that
\[
N(j+1)\geq \lambda^{2n}\left(\frac{N}{g_{\sigma(j)}^{2}P_{\sigma(j)}+N}\right)^{n}N(j).
\]
Then, by similar approach used in Theorem 1 in \cite{xu2017mean2}, the necessity of the condition \eqref{necsufMarkov} can be easily proved.
\endprf
Next, by exploiting the same TDMA scheduling scheme with the constraint
\eqref{eq:constraint} described in Section \ref{sub:Vector-systems iid},
the following result provides a necessary and sufficient condition
for the stabilizability of the vector plant \eqref{eq:vector_system}.

\begin{cor}
Given $P_{s,1},\ldots,P_{s,l},s={1,\ldots,m}$ and under the
TDMA scheduling scheme with the constraint \eqref{eq:constraint}, the process \eqref{eq:vector_system}
is mean-square stabilizable across the  AWGN fading channel  if and only if 
\begin{equation}
\prod_{i=1}^{l}|\lambda_{i}|^{\frac{2n}{l}}\rho\left(Q^{T}D\right)<1 
\label{necsufMarkov-2}
\end{equation}
with $D=\text{diag}\left(\displaystyle{\prod_{i=1}^{l}}\textstyle\left(\frac{N}{g_{1}^{2}P_{1,i}+N}\right)^{\frac{n}{l^{2}}},\ldots,\displaystyle{\prod_{i=1}^{l}}\textstyle\left(\frac{N}{g_{m}^{2}P_{m,i}+N}\right)^{\frac{n}{l^{2}}}\right).$
\end{cor}
\startprf
This result can be obtained via combining the proof of Theorem \ref{thm:2}
and Corollary \ref{cor:1}. 
\endprf
By letting the transition  matrix $Q$ be composed of $m$ identical
row vectors, which represents the probability distribution of   i.i.d process,  the results in this
section can be reduced exactly to the results in Section \ref{sec:3}
since the matrix $Q^{T}D$ in \eqref{necsufMarkov} and \eqref{necsufMarkov-2}
are rank-one matrices whose only nonzero eigenvalue equal to $\mathbb{E}_{\sigma(j)}\left[\left(\frac{N}{g_{\sigma(j)}^{2}P_{\sigma(j)}+N}\right)^{n}\right]$
and $\mathbb{E}_{\sigma(j)}\left[\prod_{i=1}^{l}\left(\frac{N}{g_{\sigma(j)}^{2}P_{\sigma(j),i}+N}\right)^{\frac{n}{l^{2}}}\right]$,
respectively.

\section{Minimum average transmit power}\label{sec:5}

In this section, we show how to derive the minimum average power $P^{*}$
satisfying the conditions proposed in  Section \ref{sec:4} via convex
optimization problems. 

Since the Markov chain $\sigma(j)$ is irreducible with all its states
positive recurrent, we have that $\sigma(j)$ has a unique stationary
distribution, denoted by the vector $\pi$. Let $R_{t}(s)$ denote
the number of visits to channel state $s$ up to time $t$, and therefore
the average occupation time of state $s$ up to time $t$ is $\frac{R{}_{t}(s)}{t}$.
Due to the fact that $\frac{R_{t}(s)}{t}\rightarrow\pi_{s}$ as $t\rightarrow\infty$,
 the average  power is given by $P=\sum_{s=1}^{m}\pi_{s}P_{s}.$

Let us start with the case of scalar plants. It should be observed
that minimizing $P$ subject to the constraint \eqref{necsufMarkov}
is a difficult problem given the complexity of the structure in the
constraint \eqref{necsufMarkov}. The following result provides an
equivalent condition for stabilizability of the process \eqref{eq:scalar system}. 

\begin{lem}
\label{lem:2}The following statements are equivalent:

1. The process \eqref{eq:scalar system} is mean-square stabilizable
across the AWGN fading channel subject to the Markov chain $\sigma(j)$.

2. $\lambda^{2n}\rho\left(Q^{T}D\right)<1$ where $D$ is
defined in Corollary \ref{cor:1}.

3. There exist $V_{s}>0,s=1,\ldots,m$ such that 
\begin{equation}\label{eq:lyapunov_cond}
V_{s}-\sum_{r=1}^{m}q_{rs}V_{r}\lambda^{2n}\left(\frac{N}{g_{s}^{2}P_{s}+N}\right)^{n}>0,s=1,\ldots,m.
\end{equation}
\end{lem}
\startprf
1$\Leftrightarrow$2 is obtained from Corollary \ref{cor:1}.  
Let us prove $2\Leftrightarrow3$. First, it can be inferred from the sufficiency proof of Corollary \ref{cor:1} that the condition $\lambda^{2n}\rho\left(Q^{T}D\right)<1$ is necessary and sufficient for  mean-square stability of the Markov jump linear scalar system $
S(j+1)=\lambda^{n}\left(\textstyle{\frac{N}{g_{\sigma(j)}^{2}P_{\sigma(j)}+N}}\right)^{\frac{n}{2}}S(j)
$
where the sequence $\sigma(j),\ j=1,2,\ldots$ is a Markov chain. Then, according to Theorem 2 in \cite{costa1993stability}, it is obtained that the above scalar system is mean-square stable if and only if the condition in \eqref{eq:lyapunov_cond} is satisfied. 
\endprf

The following result provides a solution to derive the minimum average power $P^{*}$ via a geometric program. 
\begin{thm}
\label{thm:4}Let $\bar{P}_{s}^{*},s=1,\ldots m$ be the optimal solution
of the geometric program
\begin{equation}
\begin{array}{l}
\underset{\bar{P}_{s}>0,V_{s}>0,s=1,\ldots,m}{\text{inf}}\sum_{s=1}^{m}\frac{\pi_{s}}{g_{s}^{2}}\bar{P}_{s}\\
\ \ \ \ \text{s.t.}\left\{ \begin{array}{c}
\lambda^{2n}N^{n}\bar{P}_{s}^{-n}V_{s}^{-1}\left({\textstyle \sum_{r=1}^{m}}q_{rs}V_{r}\right)<1\\
N\bar{P}_{s}^{-1}\leq1,\ s=1,\ldots,m.
\end{array}\right.
\end{array}\label{eq:GP1}
\end{equation}
Then, the minimum average  power $P^{*}$ required to stabilize the
process \eqref{eq:scalar system} across the  AWGN fading channel subject
to the Markov chain $\sigma(j)$ is given by $P^{*}=\sum_{s=1}^{m}\pi_{s}\left(\frac{\bar{P}_{s}^{*}-N}{g_{s}^{2}}\right)$.\footnote{
The actual average transmit power needed to stabilize the process should be $P^*+\epsilon$ with any $\epsilon>0$ since the feasible set in \eqref{eq:GP1} is  not compact.}\end{thm}
\startprf
First, based on Lemma \ref{lem:2}, the minimum average  power required
to stabilize the process \eqref{eq:scalar system} is given
by 
\[
\begin{array}{rl}
P^{*} & =\underset{P_{s},V_{s},s=1,\ldots,m}{\text{inf}}\sum_{s=1}^{m}\pi_{s}P_{s}\\
 & \text{s.t.}\left\{ \begin{array}{l}
V_{s}-{\textstyle \sum_{r=1}^{m}}q_{rs}V_{r}\lambda^{2n}\left(\frac{N}{g_{s}^{2}P_{s}+N}\right)^{n}>0\\
P_{s}\ge0,V_{s}>0,s=1,\ldots,m.
\end{array}\right.
\end{array}
\]

Next, by letting $\bar{P}_{s}=g_{s}^{2}P_{s}+N,s=1,\ldots,m$, the above
optimization problem can be equivalently rewritten as 
\[
\begin{array}{rl}
\underset{\bar{P}_{s},V_{s},s=1,\ldots,m}{\text{inf}}\  & \sum_{s=1}^{m}\pi_{s}\left(\frac{\bar{P}_{s}-N}{g_{s}^{2}}\right)\\
\text{s.t.} & \left\{ \begin{array}{c}
V_{s}-{\textstyle \sum_{r=1}^{m}}q_{rs}V_{r}\lambda^{2n}\textstyle{\left(\frac{N}{\bar{P}_{s}}\right)}^{n}>0\\
\bar{P}_{s}\ge N,V_{s}>0,s=1,\ldots,m.
\end{array}\right.
\end{array}
\]
Since the value of $\sum_{s=1}^{m}\pi_{s}\frac{N}{g_{s}^{2}}$ is
a  constant, it can be removed from the objective function without
affecting the optimal solution of the above optimization problem.
Thus, the above optimization problem can equivalently solved by the
geometric optimization problem \eqref{eq:GP1}. 
\endprf

For the special case where the sequence $\sigma(j),j=0,1,\ldots$
is switching according to an i.i.d. process, we have $\pi_{s}=\text{prob} (\sigma(j)=s)$.
It follows by the variable substitution $\bar{P}_{s}=g_{s}^{2}P_{s}+N,s=1,\ldots,m$
that the minimum  $P^{*}$ required to stabilize the process \eqref{eq:scalar system}
can be obtained by solving the following geometric program:
\begin{equation}
\begin{array}{rl}
\underset{\bar{P}_{s}>0,s=1,\ldots,m}{\text{inf}} & \sum_{s=1}^{m}\frac{\pi _{s}}{g_{s}^{2}}\bar{P}_{s}\\
\text{s.t.} & \left\{ \begin{array}{c}
\lambda^{2n}N^{n}\sum_{s=1}^{m}\pi_{s}\bar{P}_{s}^{-n}<1\\
N\bar{P}_{s}^{-1}\leq1,s=1,\ldots,m.
\end{array}\right.
\end{array}\label{eq:GP2}
\end{equation}
The following example shows the effect of power adaptation. 
\begin{exmp}\label{ex:2}
Consider the system described in Example \ref{ex:1} with $\lambda=1.5,N=1,n=20,g_1=1,g_2=0.5$. Let $\pi _{1}=\text{prob}(\sigma(j)=1)$ and $\pi_{2}=1-\pi_{1}.$ By solving the geometric program \eqref{eq:GP2}, we obtain the result of $P^{*}$ with different value of $\pi_{1}\in[0,1]$ plotted in Figure \ref{fig:1}. Meanwhile,
we also compute $P^{*}$ when the transmit power remains constant
with different channel state, i.e., $P_{1}=P_{2}$. As shown in Figure \ref{fig:1}, by allowing the transmit power adaptation, we can obtain a smaller
minimum average transmit power $P^{*}$.
\begin{figure}
\center \includegraphics[scale=0.26]{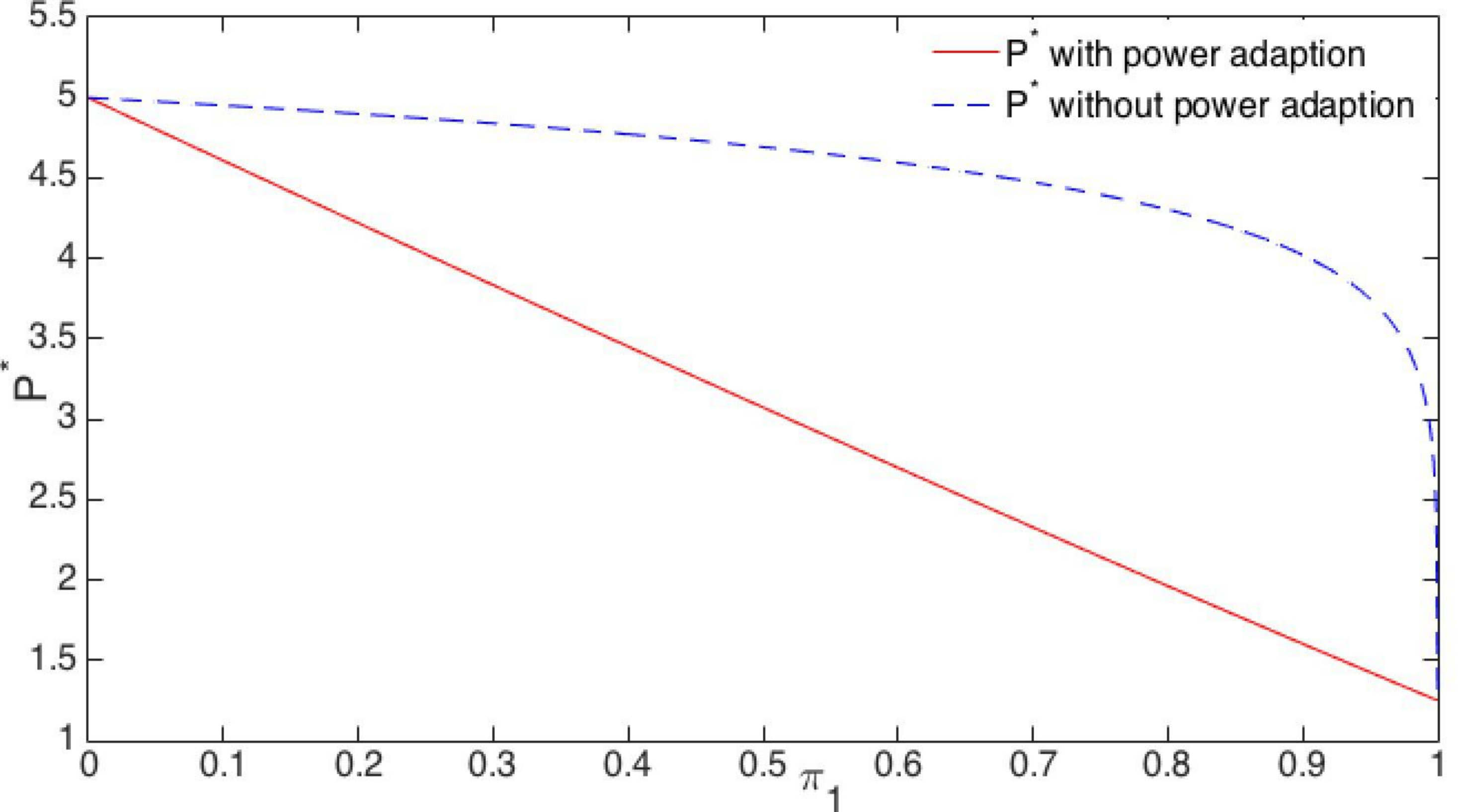}
\caption{Comparison of minimum $P^{*}$ with and without power adaptation}\label{fig:1}
\end{figure}
\end{exmp}
In the end, let us consider the vector plants. Under the TDMA scheduling
scheme with the constraint \eqref{eq:constraint}, we have that the
average transmit power is given by $P=\sum_{s=1}^{m}\sum_{i=1}^{l}\frac{\pi_{s}}{l}P_{s,i}$.
For $s=1,\ldots,m,i=1,\ldots,l$, define $\bar{P}_{s,i}=g_{s}^{2}P_{s,i}+N,$
and then the constraint \eqref{eq:constraint} can be equivalently
expressed as $\frac{|\lambda_{1}|^{2l}}{|\lambda_{i}|^{2l}}\bar{P}_{s,i}\bar{P}_{s,1}^{-1}=1$.
By replacing $\lambda^{2n}$ and $\left(\frac{N}{g_{s}^{2}P_{s}+N}\right)^{n}$
with $\prod_{i=1}^{l}|\lambda_{i}|^{\frac{2n}{l}}$ and $\prod_{i=1}^{l}\left(\frac{N}{g_{s}^{2}P_{s,i}+N}\right)^{\frac{n}{l^{2}}}$ in Lemma \ref{lem:2},
the result in Theorem \ref{thm:4} can be generalized to the case
of vector plants, which is presented as follows. 

\begin{cor}
Let $\bar{P}_{s,i}^{*},s=1,\ldots m,i=1,\ldots,l$ be the optimal solution
of the geometric program 
\begin{equation}
\begin{array}{l}
\underset{\bar{P}_{s,i}>0,V_{s>0}}{\text{inf}}{\displaystyle {\textstyle \sum_{s=1}^{m}\sum_{i=1}^{l}}}\frac{\pi_{s}}{lg_{s}^{2}}\bar{P}_{s,i}^{*}\\
\text{s.t.}\left\{ \begin{array}{c}
|\lambda_{1}|^{2n}N^{\frac{n}{l}}\bar{P}_{s,1}^{-\frac{n}{l}}V_{s}^{-1}\left({\textstyle \sum_{r=1}^{m}}q_{rs}V_{r}\right)<1\\
N\bar{P}_{s,i}^{-1}\leq1,\ \frac{|\lambda_{1}|^{2l}}{|\lambda_{i}|^{2l}}\bar{P}_{s,i}\bar{P}_{s,1}^{-1}=1\\
s=1,\ldots,m,i=1,\ldots,l
\end{array}\right.
\end{array}\label{eq:GP1-1}
\end{equation}
Then, under the TDMA scheduling scheme with the constraint \eqref{eq:constraint},
the minimum average  power $P^{*}$ required to stabilize the process
\eqref{eq:vector_system} across the  AWGN fading channel subject to
the Markov chain $\sigma(j)$ is given by $P^{*}=\sum_{s=1}^{m}\sum_{i=1}^{l}\frac{\pi_{s}(\bar{P}_{s,i}^{*}-N)}{lg_{s}^{2}}$.
\end{cor}

Similarly, the condition in \eqref{eq:GP1-1} can be reduced to a
simpler form when $\sigma(j)$ is switching according to an i.i.d.
process.

It is worth mentioning that the geometric optimization problem can
be turned into convex optimization problem via variable transformation,
see  \cite{boyd2004convex} for more details.

\section{Conclusion}
 We have investigated the minimum average transmit power required for mean-square
stabilization of a discrete-time linear process across a time-varying AWGN fading channel. Based on the  necessary and sufficient conditions for the mean-square
stabilizability we have derived, it is shown that
 the minimum required average transmit power via optimal power
adaptation can be obtained by solving a geometric program.

\bibliographystyle{IEEEtran}
\bibliography{IEEEabrv,bibbysu}

\end{document}